\documentclass[aps,pra,twocolumn,a4paper,noeprint]{revtex4-1}
\usepackage[utf8]{inputenc}
\usepackage{graphicx}
\usepackage{amsmath}
\usepackage{amsfonts}
\usepackage{amssymb}
\usepackage{bbold}
\usepackage{physics}
\usepackage[usenames,dvipsnames,svgnames,table]{xcolor}
\usepackage[USenglish]{babel}
\usepackage[colorlinks,linkcolor=blue,urlcolor=blue,citecolor=blue]{hyperref}
\usepackage{grffile}
\usepackage{float}

\newcommand{\mytitle}{Electric field induced second-order anomalous Hall transport in unconventional Rashba system}

\begin{document}

\title{\mytitle}

\author{Ankita Bhattacharya}
\email{ankita.bhattacharya@physics.uu.se}
\affiliation{Department of Physics and Astronomy, Uppsala University, Box 516, S-751 20 Uppsala, Sweden}
\author{Annica M. Black-Schaffer}
\email{annica.black-schaffer@physics.uu.se}
\affiliation{Department of Physics and Astronomy, Uppsala University, Box 516, S-751 20 Uppsala, Sweden}
\date{\today}
\begin{abstract}
 Nonlinear responses in transport experiments may unveil information and generate new phenomena in materials that are not accessible at linear order due to symmetry constraints. While the linear anomalous Hall response strictly requires the absence of time-reversal symmetry, the second-order, thus nonlinear, Hall response needs broken inversion symmetry. Recently, much effort has been made to obtain a second-order Hall voltage in response to a longitudinal ac driving current, both to obtain information about band geometric quantities and for its useful technological applications, including rectification and frequency doubling. Typically, additional material engineering is required in noncentrosymmetric systems to obtain second-order responses since it obeys a stringent crystallographic symmetry constraint. To circumvent this, an alternative route is to apply a dc electric field. In this Letter, we uncover an electric field induced second-order anomalous Hall effect in an inversion-broken system possessing unconventional Rashba bands. We establish that the quantum metric, a geometrical feature of electronic wave functions providing information on non-trivial structure of Bloch bands, is responsible for providing the nonlinear Hall response. 
\end{abstract}

\maketitle

{\color{blue}{\textit{Introduction.}}}---Recent times have witnessed a surge of interest in the quantum geometry of electrons, due to the pivotal role it has been shown to play in condensed matter \cite{Torma23,NS10, Resta11,Peotta15,Ozawa21}, quantum information \cite{Hou24}, optics \cite{AGN22, Gao19}, and statistical physics \cite{BC94}. As a key example, the quantum geometric tensor\,(QGT) \cite{Kibble79,Provost80,Marzari97,Brody2001,Cheng10,Torma23}, defined in a parameterized Hilbert space and determining the geometry of eigenstate space, has been shown to play an indispensable role in transport properties of materials, both through its real part labeled as the quantum metric (QM) and its imaginary part, the Berry curvature (BC). Besides determining the topological characteristics of quantum materials, the BC emerges in various Hall effects \cite{XCN10, NS10, TKN82}, including leading to a linear anomalous Hall effect in solids with broken time-reversal symmetry\,(TRS) without the need of any external magnetic fields \cite{NS10}. At the same time, the QM has been shown to be essential in nonlinear transport theory \cite{Gao14}.
Furthermore, the QM has recently also been established as a key ingredient for superfluidity in flat band systems \cite{Peotta15,Bernevig20,TPB22}, fractional Chern insulators \cite{Roy14,Bergholtz22,Bergholtz24}, Lamb shift in transitional-metal dichalcogenides \cite{Srivastava15}, and orbital magnetism \cite{GYN15, PRFM16}. 

Recently, Sodemann and Fu \cite{SoFu15} proposed a second-order anomalous Hall effect in \emph{noncentrosymmetric time-reversal-invariant} systems where instead of the BC, the dipole moment of BC over the occupied states, known as the Berry curvature dipole (BCD) is the key ingredient. Subsequently, BCD-assisted nonlinear transport have been  detected experimentally in several nonmagnetic materials \cite{MXS19, Herrero18,KLS19,HIZ21,HWH22,ZCZ23}. However, a major caveat is that the BCD follows a strict symmetry constraint: the maximum symmetry allowed is the presence of a single mirror line in the transport plane \cite{SoFu15, Ortrix21}. This means the BCD is still zero in many high-symmetry but still inversion-broken systems. In these cases additional material engineering, such as interlayer twisting \cite{HCH21}, applying strain \cite{Araki18,Son19,Qin21, Narayan23} or an electric field \cite{Ye23}, is required to lift the symmetry constraint to observe BCD-assisted Hall effect. In particular, using an electric field utilizes the presence of another band geometric quantity, the Berry connection polarizability (BCP) \cite{Gao14,Lai21,Liu22}, which is also the so-called band-normalized QM \cite{Ortrix24}. This intrinsic band geometric quantity in the presence of an electric field  provides a field-induced BCD, even in systems where the native BCD (and BC) are absent by symmetry \cite{Ye23}. An additional experimental advantage is that the field-induced BCD is linearly proportional to the applied electric field, and therefore, the Hall response can easily be tuned externally.

The simplest inversion-broken system that preserves TRS is likely the 2D electron gas with Rashba spin-orbit coupling (RSOC), which not only is of fundamental importance in condensed matter \cite{SG21,Rashba22} but also highly relevant for spintronics applications \cite{DS04}. Due to structural inversion asymmetry, spin degeneracy is removed in the presence of RSOC, without requiring TRS breaking. This usually leads, with only moderate doping level tuning, to two concentric split Fermi circles having opposite spin chirality \cite{SG21}. However, in the presence of additional degrees of freedom, e.g., orbitals, it has also been reported that the spin chirality of the two spin split Fermi surfaces can be the same. Such bands have recently been coined as \emph{unconventional} Rashba bands \cite{Huang24}, and have been reported in at least monolayer $\mathrm{Os}\mathrm{Bi}_2$ \cite{SHZ21}, in the surface alloy of $\mathrm{Bi}/\mathrm{Cu}(111)$ \cite{MHE09}, $\mathrm{Ag}-\mathrm{Au}(111)$ \cite{Bentmann12,NK19}, and in trigonal layered $\mathrm{Pt}\mathrm{Bi}_2$ \cite{GZZ18}.

In this Letter, we study field-induced and BCD-assisted nonlinear, to second order, anomalous Hall response in a 2D unconventional Rashba system using a simple prototype model \cite{Huang24}.
Similar to conventional Rashba systems, we find that the inherent BCD also vanishes in unconventional Rasbha systems. 
Further, conventional Rashba systems do have a finite BCP, or equivalently QM, due to its spin-momentum locked band structure. Still, electric field-induced transport is not feasible in conventional Rashba systems due to the fact that the QM-induced contributions of the two bands exactly cancel each other at every momenta \cite{Ortrix24}. Either hexagonal band warping  \cite{Saha23,Pal24} or an external magnetic field \cite{Ortrix24} is required to observe nonlinear transport in conventional Rashba systems. In contrast, we find a highly tunable \emph{second-harmonic} Hall response induced by electric field in a unconventional Rashba system, due to its non-trivial QM. This opens up for nonlinear Hall transport in several spin-orbit coupled materials without breaking TRS and without requiring additional materials engineering.

\begin{figure}
    \centering
   \includegraphics[width=0.80\columnwidth]{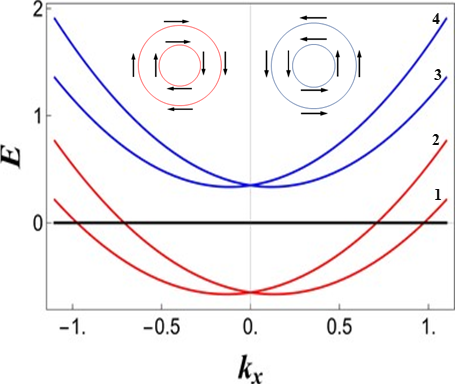} 
  \caption{Energy dispersion of unconventional Rasbha system in Eq.~(\ref{model}). Spin chiralities of the lower red bands (marked as $1$ and $2$) are the same but opposite to the upper blue bands (marked as $3$ and $4$), following from Eq.~(\ref{Sx}) and Eq.~(\ref{Sy}) and schematically shown in the insets. Parameters are set to $t=1$, $\lambda=0.6$, $\eta=0.8$, and $\mu=0.65$ such that only the red bands cross the Fermi energy (black line).}
    \label{Bs}
\end{figure}

{\color{blue}{\textit{Unconventional Rashba system.}}}---We start by introducing a generic, yet simple, model in $2$D for unconventional Rashba systems. For this we employ the $\mathbf{k}\cdot\mathbf{p}$ Hamiltonian  that encodes the unconventional Rashba bands \cite{Huang24,Wang24}:
\begin{align}
    H(\mathbf{k}) = h_0 \,\tau_0 \sigma_0 - \lambda\,(\tau_0 + \tau_1)\,(\,k_y\,\sigma_x - k_x \sigma_y)+ \eta\,\tau_2\,\sigma_3.
    \label{model}
\end{align}
Here we use the basis $\psi_{\mathbf{k}}=(c_{\mathbf{k},1,\sigma},c_{\mathbf{k},2,\sigma})^T$, 
where $\sigma=\uparrow, \downarrow$ describes spin and $j=1,2$ corresponds to orbitals, while $\tau_i$ and $\sigma_i$ denote the Pauli matrices for $i={1,2,3}$ and the $2\times 2$ identity matrix for $i=0$, spanning the orbital and the spin spaces, respectively. Further, $h_0=(t k^2-\mu)$, where the parameter $t$ controls the band curvature and $\mu$ is the chemical potential, while
$\lambda$ is the strength of the $k$-linear RSOC and $\eta$ gives an on-site spin-orbit coupling (SOC).
The nearest-neighbor $\lambda$ term, which breaks the inversion symmetry, couples the electrons of opposite spins from both the same and different orbitals, while the on-site $\eta$ term describes the interactions between electrons of different orbitals within same site and is responsible for splitting of the band structure at the $k=0$ point. The eigenenergies of Eq.~(\ref{model}) are $\xi_{s_1,s_2}(\mathbf{k})=h_0 + s_1 \sqrt{\eta^2 + k^2 \lambda^2} + s_2\, k \lambda$, where $s_1,s_2 \in \{+,-\}$, and $k= \sqrt{k_x^2 + k_y^2}$. The dispersion is thus isotropic, i.e., independent of $\theta= \arctan{(k_y/k_x)}$. We plot the dispersion in Fig.\,\ref{Bs}. The spin texture for the lowest two bands, called $1$ and $2$ and red in Fig.\,\ref{Bs} is
\begin{align}
    (S^{1,2}_x, S^{1,2}_y) = \frac{k \lambda}{\sqrt{k^2 \lambda^2 + \eta^2}}\, (\sin{\theta}\, \hat{e}_x - \cos{\theta}\,\hat{e}_y), 
    \label{Sx}
\end{align}
while for the bands $3$ and $4$  i.e., two blue bands in Fig.~\ref{Bs}, the spin texture is opposite
\begin{align}
    (S^{3,4}_x, S^{3,4}_y) = \frac{k \lambda}{\sqrt{k^2 \lambda^2 + \eta^2}}\, (-\sin{\theta}\, \hat{e}_x + \cos{\theta}\,\hat{e}_y),
    \label{Sy}
\end{align}
where $S_i= \bra \psi  \tau_0 \bigotimes \sigma_i \ket \psi$, with $\psi$ being the eigenstate of the Hamiltonian in Eq.~(\ref{model}). As a consequence, two Fermi surfaces of same spin chirality are obtained when $\mu$ is tuned such that only the red bands cross the Fermi energy. This is in contrast to conventional Rashba bands where, at any filling, two Fermi surfaces of opposite spin chiralities are instead obtained.

{\color{blue}{\textit{Theory of second-order anomalous Hall effect.}}}---We next briefly review the theory of second-order anomalous Hall response in order later to apply it to the unconventional Rashba system. Within the framework of semiclassical Boltzmann formalism, the current density $j_\gamma$ along direction $\gamma$ in dimensions $d$ can be expressed as the integral of the product of the velocity of electrons $v_\gamma$ and the non-equilibrium occupation function $f(\mathbf{k})$
\begin{align}
    j_\gamma = -e \int \frac{d^d\mathbf{k}}{(2\pi)^d}\, f(\mathbf{k})\, v_\gamma
    \label{current}
\end{align}
To obtain the current response up to second-order in electric field, correction up to first order in electric field is considered both in $f(\mathbf{k})$ and $v_\gamma$.
In the steady-state, under the relaxation time approximation, $f(\mathbf{k})$ is expanded up to first order in electric field $\boldsymbol{\mathcal{E}}$ as \cite{Mahan80}
\begin{align}
    f^{(1)}(\mathbf{k})= f_0 (\mathbf{k}) + \frac{e\tau}{\hbar}\, \boldsymbol{\mathcal{E}} \cdot (\nabla_{\mathbf{k}} f_0),
\end{align}
where $f_0=\frac{1}{ 1\, +\, e^{\beta (\epsilon_{\mathbf{k}}-\mu)}}$ is the equilibrium Fermi-Dirac distribution function with $\beta=1/k_B T$ the inverse temperature. Here, $\tau$ is the relaxation time. 
Further, in the presence of a nonzero BC, electrons acquire an additional anomalous velocity linear in the applied electric field
$\mathbf{v}_\text{an} = \frac{e}{\hbar} \boldsymbol{\mathcal{E}}\times \mathbf{\Omega}_n$, where $\mathbf{\Omega}_n$ is the BC of $n$-th band. It is evident that $\mathbf{v}_\text{an}$ is necessarily transverse to the applied field. 
Since  we are interested in the Hall response, we in the following consider only the transverse component of the resulting current.
Substituting the expressions for $f^{(1)}(\mathbf{k})$ and the anomalous velocity into Eq.~(\ref{current}), we thus obtain the anomalous Hall current response up to second order in electric field as
\begin{align}
    j^{\text{AH}}_{\mu}=\sigma^{\text{AH}}_{\mu \nu} \,\mathcal{E}_\nu + \chi^{\text{AH}}_{\mu \nu \lambda}\, \mathcal{E}_{\nu} \mathcal{E}_{\lambda}.
\end{align}
The transverse conductivity term $\sigma^{AH}_{\mu \nu}=e/\hbar\sum_{n} \int \frac{d^d \mathbf{k}}{(2 \pi)^d}\, f_0 \,\Omega_{n \delta}$, yields the linear anomalous Hall response. This term vanishes in the presence of TRS owing to the fact that under time-reversal operation\,($\mathcal{T}$) BC transforms as an odd function, $\Omega_n (\mathbf{k})\xrightarrow{\mathcal{T}} -\Omega_n (-\mathbf{k})$ \cite{XCN10}, while $f_0 (\mathbf{k}) \xrightarrow{\mathcal{T}} f_0 (\mathbf{-k})$ thus making the momentum integration vanishing. 
However, the second-order transverse conductivity tensor $\chi^H_{\mu \nu \lambda}$ may survive in the presence of TRS, which is the focus of this work. 

To proceed, for an ac driving electric field $\mathcal{E}^\omega$ with frequency $\omega$, a second-harmonic transverse current $j^{2\omega}_\mu = \chi^{\text{AH}}_{\mu \nu \lambda} \mathcal{E}^\omega_\nu \mathcal{E}^\omega_\lambda$ appears i.e.~with frequency twice the driving frequency, in addition to a dc rectified current $j^0_\mu = \chi^{\text{AH}}_{\mu \nu \lambda} \mathcal{E}^\omega_\nu {\mathcal{E}^\omega_\lambda}^*$. For such ac driving field, the second-order transverse conductivity tensor has been shown to simplified to \cite{SoFu15}
\begin{align}
   \chi^{\text{AH}}_{\mu \nu \lambda} = \epsilon_{\mu \delta \lambda}\, \frac{e^3 \tau}{2\hbar^2 (1 + i \omega \tau)}\, \sum_{n}\, \int \frac{d^d \mathbf{k}}{(2 \pi)^d}\, f_0 \,(\partial_{k_\nu} \Omega_{n\delta}),
    \label{chi_def}
\end{align}
where $\epsilon_{\mu \delta \lambda}$ is the Levi-Civita symbol. The last part is the integral of the first moment of the BC over the occupied states, also called the BCD \cite{SoFu15}:
\begin{align}
    D_{\nu \delta} = \sum_{n}\,\int \frac{d^d \mathbf{k}}{(2 \pi)^d}\, f_0 \,(\partial_{k_\nu} \Omega_{n\delta}).
    \label{BCD_def}
\end{align}
Interestingly, $\chi^{\text{AH}}_{\mu \nu \lambda}$ survives in the presence of TRS since the momentum integration in Eq.~(\ref{chi_def}) is non-vanishing following the derivative $\partial_{k_\nu} \Omega_{n}(\mathbf{k})$ being even under $\mathcal{T}$. However, inversion symmetry must be broken in order to achieve a nonzero $\chi^{\text{AH}}_{\mu \nu \lambda}$, since otherwise $ \Omega_{n}(\mathbf{k})\xrightarrow{\mathcal{P}} \Omega_{n}(-\mathbf{k} )$ would make the derivative $\partial_{k_\nu} \Omega_{n}(\mathbf{k})$ an odd function of momentum under the inversion operation $\mathcal{P}$. In addition, the BCD obeys a severe crystal symmetry constraint that we discuss in the next section. Before proceeding, we note that $\chi^{\text{AH}}_{\mu \nu \lambda}$ also depends on the quasiparticle scattering time $\tau$, making it non-intrinsic. However, at frequencies below the interband threshold but above the width of the Drude peak $\omega \tau \gg 1$, the prefactor in Eq.~(\ref{chi_def}) becomes independent of $\tau$~\cite{SoFu15,Ortrix21}. We note that at second order, such an intrinsic transverse response not depending on $\tau$ is also obtainable if TRS is broken~\cite{Liu21,Das23}. In fact, in a $\mathcal{P}\mathcal{T}$-invariant antiferromagnet this intrinsic contribution has been shown to be dominant~\cite{Liu21}.

{\color{blue}{\textit{Symmetry of $\chi^{\text{AH}}$.}}}---We next focus on the symmetry properties of second-order conductivity tensor $\chi^{\text{AH}}$. As discussed above,  $\chi^{\text{AH}}$ survives in TRS-invariant systems, provided that inversion symmetry is broken. To establish further symmetry properties we note that in $2$D the BC has only one (out of the plane) component $\Omega_{n z}$, when assuming $x-y$ to be the transport plane, which makes it a pseudoscalar \cite{NS10}. 
As a consequence, the BCD in Eq.~(\ref{BCD_def}) has only two relevant components, $\{D_{xz}, D_{yz}\}$ and thus, acts as pseudovector in $2$D. Further, from Eq.~(\ref{chi_def}) it is evident that $\chi^{\text{AH}}$ is directly proportional to the BCD and therefore, also it transforms as a pseudovector in $2$D.
 Therefore, under any point group operation, which can be represented by an orthogonal matrix $R$, the pseudovector $\tilde{\chi}^{\text{AH}}$ transforms as
\begin{align}
    \tilde{\chi}^{\text{AH}}_{\beta}= \det{R}\, R_{\beta \alpha}\, \tilde{\chi}^{\text{AH}}_{\alpha}.
    \label{chi_symm}
\end{align}
This transformation rule indicates that the presence of two or more mirror lines in a crystal forces the second-order Hall response to vanish identically \cite{SoFu15,Ortrix21,NaFu19}. On the other hand, a single mirror line forces it to be orthogonal to that mirror plane. This is quite a stringent symmetry constraint on the BCD, rendering it zero in many high-symmetry but still noncentrosymmetric materials.

In both conventional and unconventional Rashba systems the BC is not only zero, but it also has two mirror lines, $\mathcal{M}_x$ ($\tau_x\,\sigma_x$) and $\mathcal{M}_y$ ($\tau_y\,\sigma_y$), which also forces the BCD to vanish. As such the prospects of a finite Hall response seems dim. However, a recently discovered route to obtain a finite anomalous Hall response in a system where the intrinsic BCD vanishes is to apply a dc electric field to induce a (dc) field-induced BC, which may also lead to a field-induced BCD due to the fact that the applied dc field lowers the symmetry of the system \cite{Ye23}. This is the possibility we discuss next.

{\color{blue}{\textit{Electric field induced Hall effect.}}}---
In the presence of a static electric field, a field-induced BC $\mathbf{\Omega}^{E}_n = \nabla \times \mathcal{A}^{E}_n$ is generated, attributed to the positional shift of the electron wave packets under application of electric field~\cite{Gao14}. The field-induced Berry connection can be expressed as $\mathcal{A}^{E}_{n,a}= G^n_{ab} E_b$, with $G^n_{ab}$ being the BCP defined as \cite{Gao14,Lai21,Liu22}
\begin{align}
   G^n_{ab}= 2 \Re \sum_{m\neq n} \frac{ \bra{u_n} i \partial_{k_a} \ket{u_m}\bra{u_m} i \partial_{k_b} \ket{u_n}}{\epsilon_n-\epsilon_m}.
    \label{BCP_def}
\end{align}
Here, $\ket{u_m}$ is the periodic part of the $m$-th Bloch eigenstate, while $\epsilon_m$ denotes the eigenenergy of $m$-th band. We note that $G^n_{ab}$ is concomitant to the band-resolved QM, $g^n_{ab}= 2 \Re\,[\bra{u_n} i \partial_{k_a} \ket{u_m}\bra{u_m} i \partial_{k_b} \ket{u_n}]$, which is the real part of the QGT \cite{Provost80,Torma23}.
Now if a dc electric field $\mathbf{E}^\text{dc}$ is applied in the $x-y$ transport plane also defining the 2D material, such that it makes an angle $\theta$ with the $x$-axis, $\mathbf{E}^{\text{dc}}= E^{\text{dc}}\, (\cos\theta, \sin\theta)$, the induced BC can be expressed as \cite{Lai21, Pal24}
\begin{align}
\label{EinBC}
  \Omega^{\text{E}}_{n z} &=E^{\text{dc}}[(\partial_{k_x} G^n_{yx}- \partial_{k_y} G^n_{xx})\,\cos\theta \\ \nonumber
     & + (\partial_{k_x} G^n_{yy}- \partial_{k_y} G^n_{xy})\,\sin\theta].
\end{align}
Since in the presence of $\mathbf{E}^\text{dc}$, the symmetry of the system is lowered this field-induced BC may also lead to a finite BCD, which then also develops an angular dependence $\mathbf{D}^{\text{E}}(\theta) = (D^{\text{E}}_{xz} (\theta),D^{\text{E}}_{yz} (\theta))$. 

Due to a field-induced BCD, a second-order anomalous Hall effect can also be induced. Assuming an in-plane ac driving field $\mathbf{E}^\omega$ to probe the Hall effect, satisfying $ E^\omega \ll E^{\text{dc}}$, we arrive at the second-harmonic Hall current being \cite{Ye23}
\begin{align}
    \mathbf{j}^{2\omega}= -\frac{e^3 \tau}{2 (1 + i \omega \tau) \hbar^2}\, (\hat{z}\times \mathbf{E}^\omega ) \,[\mathbf{D}^{\text{E}}(\theta) \cdot \mathbf{E}^\omega].
    \label{acHall2nd}
\end{align}
Note that we here rephrase Eq.~(\ref{chi_def}) in terms of the second-harmonic current ($\chi^{\text{AH}}= j^{2\omega}/(E^\omega)^2$) and make the angular dependence of the BCD explicit. The direction of the Hall current is always perpendicular to $\mathbf{E}^\omega$ as expecte for the transvers response, while its magnitude depends on its projection onto $\mathbf{D}^{\text{E}}(\theta)$. Thus the maximum Hall response is obtained when $\mathbf{E}^\omega \parallel \mathbf{D}^{\text{E}}(\theta)$, but it vanishes for $\mathbf{E}^\omega \perp \mathbf{D}^{\text{E}}(\theta)$.
\begin{figure}[t!]
    \centering
   \raisebox{1ex} {\includegraphics[width=\columnwidth]{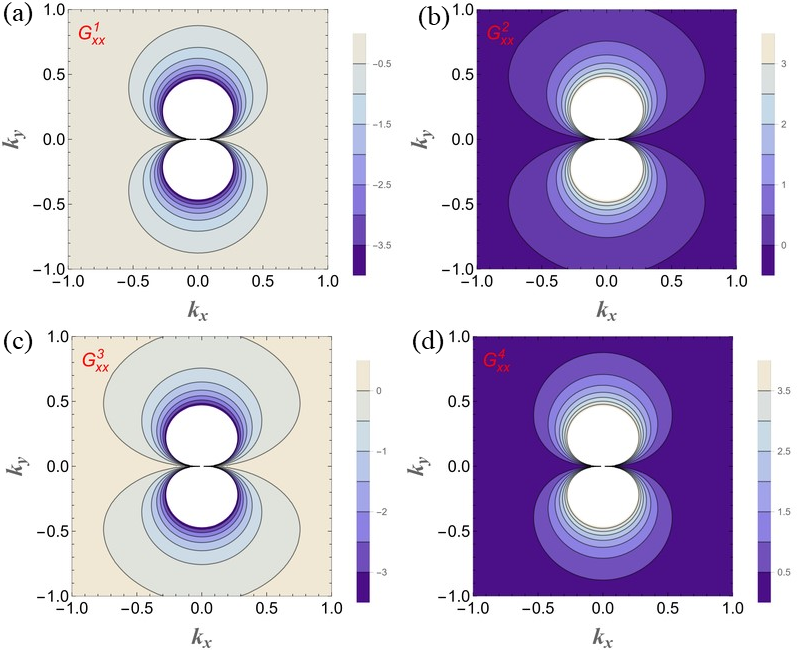} }
\caption{The $\mathbf{k}$-resolved distribution of the component $G_{xx}$ of the BCP tensor in Eq.~\eqref{BCP_def} for the four unconventional Rashba bands. Here, $1, 2$ (top row) correspond to the red bands and $3, 4$ (bottom row) correspond to the blue bands in Fig.~\ref{Bs}, respectively. Parameters are same as in Fig.~\ref{Bs}. }
    \label{BCP}
\end{figure}
\begin{figure}[b!]
    \centering
   \raisebox{1ex} {\includegraphics[width=0.85\columnwidth]{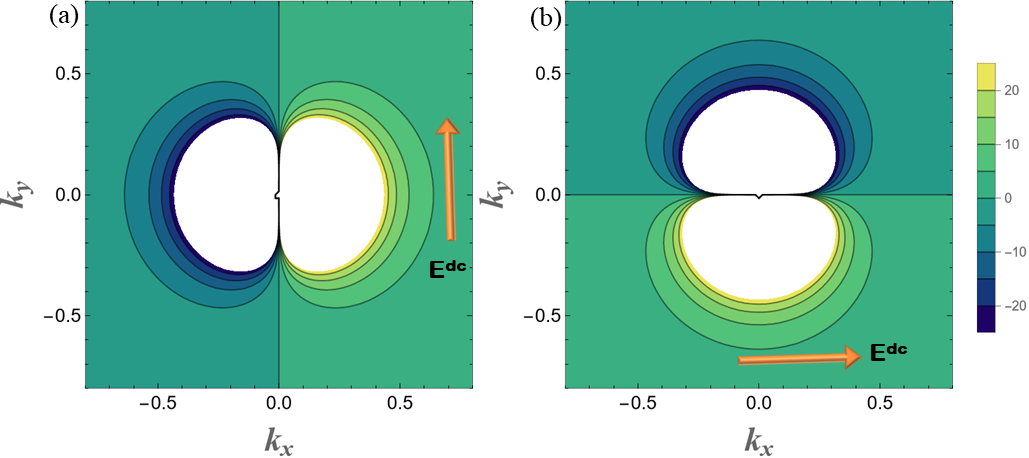}}
    \caption{The $\mathbf{k}$-resolved distribution of the electric field induced BC for band $1$ in Fig.~\ref{model} when the electric field $\mathbf{E}^{\text{dc}}$ is (a) along $y$- and (b) along $x$-direction. The BC clearly shows a dipole-like distribution that leads to finite field-induced BCD with the dipole oriented along the direction of $\mathbf{E}^{\text{dc}}$. Magnitude of $\mathbf{E}^{\text{dc}}$ is set to unity. Parameters are same as in Fig.~\ref{Bs}.}
    \label{BCE}
\end{figure}

To show how a second-order anomalous Hall effect induced by an applied electric field is generated in an unconventional Rashba system, we start by exploring the BCP in Fig.~\ref{BCP}. To showcase the BCP we plot the $\mathbf{k}$-resolved distribution of the component $G_{xx}$ of the BCP tensor defined in Eq.~(\ref{BCP_def}) for all four unconventional Rashba bands in Fig.~\ref{model}. As seen, the sum of the contributions of BCP tensor vanishes when contributions from all four bands are taken into account. However, if the chemical potential is tuned such that only a subset of bands, and particularly bands $1$ and $2$ (see Fig.~\ref{model}) cross the Fermi energy, the BCP contribution becomes non-vanishing for the occupied states. We note that this is in contrast to conventional Rashba systems, where the BCP contributions always exactly cancel.

\begin{figure}[t!]
    \raisebox{1ex} {\includegraphics[width=0.95\columnwidth]{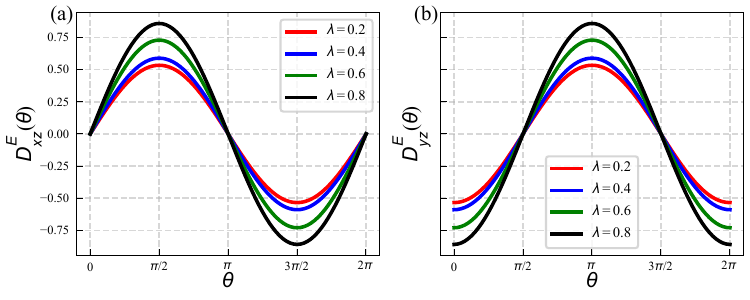}} \\
   \raisebox{1ex}{ \includegraphics[width=0.95\columnwidth]{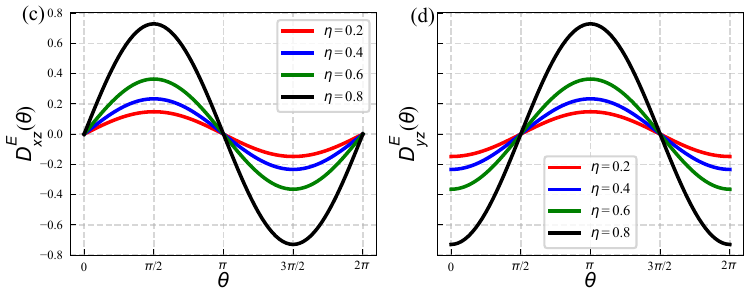} }
  \caption{Components of electric field-induced BCD as a function of electric field $\mathbf{E}^{\text{dc}}$ angle $\theta$ with the variation of RSOC term $\lambda$ (a,b) and on-site SOC $\eta$ (c,d). The $x$-component of BCD $D^{\text{E}}_{xz}$ shows $\sin \theta$-like (a,c) and its $y$-component $D^{\text{E}}_{yz}$ shows $\cos\theta$-like (b,d) dependence, respectively. Magnitude of $\mathbf{E}^\text{dc}$ is set to unity. Other parameters are same as in Fig.~\ref{Bs}}
    \label{BCD2}
\end{figure}

The finite BCP in turn provides a finite field-induced BC following Eq.~(\ref{EinBC}), as long as the external $\mathbf{E}^{\text{dc}}$ field lowers the symmetry sufficiently. We illustrate that that is indeed the case for the unconventional Rashba system in Fig.~\ref{BCE}, where we plot the $\mathbf{k}$-resolved distribution of the $\mathbf{E}^\text{dc}$-induced BC $\mathbf{\Omega}_n^E$ for two orientations of $\mathbf{E}^{\text{dc}}$. The results clearly shows a dipole-like distribution of the BC, which thus directly translates to a finite field-induced BCD. 
We also note that not only the amplitude but also the orientation of the field-induced BC, and thus the resulting BCD, is directly tuned by changing the directions of the $\mathbf{E}^{\text{dc}}$. Thus, the second-order Hall current is highly tunable by the external control of $\mathbf{E}^\text{dc}$. We display the resulting angular dependence of the field-induced BCD in Fig.~\ref{BCD2} by plotting its two components, $D^{\text{E}}_{xz}$ (a,c) and $D^{\text{E}}_{yz}$ (b,d), as a function of $\theta$ and uncovering clear $\sin\theta$ and $\cos\theta$ dependence, respectively. In Fig.~\ref{BCD2} we also vary the RSOC strength $\lambda$ (a,b) and on-site SOC $\eta$ (c,d) and both $D^{\text{E}}_{xz}$ and $D^{\text{E}}_{yz}$ increases with increasing $\lambda$ and $\eta$. We here note that at least $\lambda$ is known to be experimentally adjustable with the aid of a gate voltage \cite{CGG10,Yin20}. 

\begin{figure}[htbp]
    \centering
   \raisebox{1ex} {\includegraphics[width=\columnwidth]{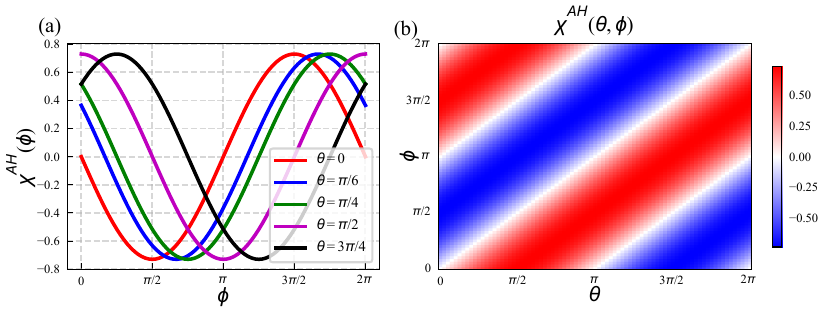} }
  \caption{(a) Second-order Hall conductivity $\chi^{\text{AH}}$ using an ac driving field $\mathbf{E}^\omega$ as a function of angle $\phi$ made by the ac driving field for various angles $\theta$ made by the symmetry breaking static field $\mathbf{E}^{\text{dc}}$ with the $x$-axis and (b) as a function of both $\theta$ and $\phi$. Parameters are same as in Fig.~\ref{Bs} and $\chi^{\text{AH}}$ is scaled by the prefactor $\mathcal{A}=\frac{e^3 \tau}{2 (1 + i \omega \tau) \hbar^2}$
  }
    \label{chi}
\end{figure}

 Having established a finite and field tunable BCD, we next apply an in-plane ac driving field $\mathbf{E}^\omega$ at an angle $\phi$ from the $x$-axis. Then $\chi^{\text{AH}}$ shows an angular dependence not only with $\theta$ due to the BCD $\mathbf{D}^{\text{E}}(\theta)$, but also with $\phi$. From Eq.~(\ref{chi_def}) it also follows that only four components of the tensor $\chi^{\text{AH}}$, namely $\chi_{yxx}$, $\chi_{xxy}$, $\chi_{xyy}$, and $\chi_{yyx}$, are non-vanishing and only two of them are also independent since  $\chi_{yxx}=-\chi_{xxy}$ and $\chi_{xyy}=-\chi_{yyx}$. Thus, for extracting the Hall current, $\chi^{\text{AH}} (\theta, \phi)$ can be simplified to \cite{Liu21}
\begin{align}
    \chi_{AH} (\theta, \phi)= \chi_{yxx} (\theta) \cos{\phi}-\chi_{xyy}(\theta) \sin{\phi}.
\end{align}
The intricate angular dependence of the second order Hall conductivity $\chi^{\text{AH}} (\theta, \phi)$ in an unconventional Rashba system is shown in Fig.~\ref{chi}. As seen, for $\mathbf{E}^\text{dc} \parallel \mathbf{E}^\omega$ the response vanishes, while it is maximum when $\mathbf{E}^\text{dc} \perp  \mathbf{E}^\omega$. This follows from two facts. First, according to Eq.~\eqref{acHall2nd} the Hall current vanishes identically if $\mathbf{D}^{\text{E}} \perp \mathbf{E}^\omega$ and is maximum for $\mathbf{D}^{\text{E}} \parallel \mathbf{E}^\omega$. Second, the field-induced $\mathbf{D}^{\text{E}}$ is always perpendicular to $\mathbf{E}^{\text{dc}}$, since in the presence of $\mathbf{E}^{\text{dc}}$ the residual mirror symmetry of the system forces the BCD to be perpendicular to the residual mirror plane following Eq.~(\ref{chi_symm}). The resulting intricate angular dependence provides extensive experimental control of the Hall current, including being able to switch it on and off easily by simply changing the relative orientation of $\mathbf{E}^\text{dc}$ and  $\mathbf{E}^\omega$.

{\color{blue}{\textit{Conclusions.}}}---We study the second-order Hall response in a time-reversal-symmetry (TRS) invariant two-dimensional (2D) system in the presence of an on-site spin-orbit coupling (SOC) that couples electrons of different orbitals within same site in addition to $k$-linear Rashba (RSOC) that couples electrons of opposite spins from same and different orbitals leading to \emph{unconventional} Rashba bands. The main dissimilarity with conventional Rashba systems is that it is possible to obtain two occupied bands with the same spin chiralities. We find that this allows for a finite Berry connection polarization (BCP), or concomitant quantum metric (QM) for the occupied states, unlike the situation for conventional Rashba systems. As a consequence, an in-plane dc electric field can generate a tunable Berry curvature dipole (BCD), which in turn directly gives a finite second-order Hall response. We probe this Hall response by applying a ac driving field in the transport plane, and thereby establish the existence of a second-harmonic Hall current in an unconventional Rashba system.
We note particularly that no additional material engineering is required to generate this second-order Hall response. Furthermore, the second-order Hall conductivity tensor shows not only periodic angular dependence with the ac driving field but also a strong angular dependence with the symmetry-reducing dc electric field. 
Since the field-induced BCD is also linearly proportional to the applied dc field, the second-order Hall current can thus be extensively tuned both by the magnitude of the applied symmetry-reducing dc field and by tuning the relative angles between the symmetry-reducing dc field and driving ac field. Such extensive tuning makes the system desirable for  technologies such as rectification or frequency doubling \cite{Isobe20}.

In summary, we report on a finite second-order Hall response in the presence of TRS in a relatively simple system due to non-trivial QM of the occupied states that should be straightforward to verify experimentally following e.g.~the experimental success in $\mathrm{WTe_2}$ \cite{Ye23}. Finally, we note that disorder plays a critical role for the nonlinear Hall effect, especially with the second-order Hall current in TRS-invariant system being purely extrinsic, disorder-induced contributions may be of importance \cite{DWL19,CDS24}. However, this is beyond the scope of this Letter and left for future work.

The authors thank Q.~Marsal, R.~Arouca, K. Das, and A.~K.~Ghosh for useful discussions related to this work. 
We acknowledge financial support from the Knut and Alice Wallenberg Foundation through the Wallenberg Academy Fellows program ~KAW 2019.0309 and project grant KAW 2019.0068 and the Swedish Research Council (Vetenskapsr\aa det) grant agreement no.~2022-03963.

\end{document}